\documentclass{aa}
\usepackage{graphicx,natbib}
\bibpunct{(}{)}{;}{a}{}{,}

\def\apj{ApJ}

\def\aap{A\&A}

\begin{document}
\sloppypar

\title{Flare magnetic reconnection and relativistic particles in
the 2003 October 28 event}

\author{C. Li \inst{1, 2}, Y. H. Tang \inst{1}, Y. Dai \inst{1}, C. Fang \inst{1}
\and J. -C. Vial \inst{2}} \institute{Department of Astronomy,
Nanjing University, Nanjing 210093, China\\
\email{chuan.li@ias.u-psud.fr; lic@nju.edu.cn} \and Institut
d'Astrophysique Spatiale, Batiment 121, Universit$\acute{\rm e}$
Paris-sud 11 and CNRS, Orsay 91405, France}
\date{Received 19 December 2006 / Accepted 11 June 2007}
\authorrunning{Li, Tang, Dai, Fang, \and Vial}
\titlerunning{Magnetic reconnection and SEPs on 28/10/03}

\abstract{An X17.2 solar flare occurred on 2003 October 28,
accompanied by multi-wavelength emissions and a high flux of
relativistic particles observed at 1 AU. We present the analytic
results of the TRACE, SOHO, RHESSI, ACE, GOES, hard X-ray (INTEGRAL
satellite), radio (Onde$\check{\rm r}$ejov radio telescope), and
neutron monitor data. It is found that the inferred magnetic
reconnection electric field correlates well with the hard X-ray,
gamma-ray, and neutron emission at the Sun. Thus the flare's
magnetic reconnection probably makes a crucial contribution to the
prompt relativistic particles, which could be detected at 1 AU.
Since the neutrons were emitted a few minutes before the injection
of protons and electrons, we propose a magnetic-field evolution
configuration to explain this delay. We do not exclude the effect of
CME-driven shock, which probably plays an important role in the
delayed gradual phase of solar energetic particles.
\keywords{acceleration of particles -- Sun: magnetic fields -- Sun:
flares}}

\maketitle

\section{Introduction}
When a high flux of relativistic solar nucleons strikes Earth's
atmosphere, the straightforward neutrons and/or the nuclear
byproducts can result in a ``ground level enhancement" (GLE).
However, the exact acceleration sources of the relativistic
particles remain enigmatic. The main controversy focuses on
acceleration occurring at shocks driven by coronal mass ejections
(CMEs) or in the active region producing flares (Kahler 1994; Reames
1999; Reames 2002; Cane et al. 2002; Kallenrode 2003). The
acceleration during the process of the coronal magnetic field
reconfiguration at heights between $0.1 R_{s}$ and $1 R_{s}$ above
the photosphere, neither in the flare active region nor at the bow
shock of the CME, is also suggested by \citet{Kle01}.

Magnetic reconnection in the active region is a candidate for a
large energy release at the Sun. When a coronal flux rope loses
equilibrium and travels upwards, below which an extensive
reconnection current sheet (RCS) forms, the reconnection in this RCS
releases most of the magnetic energy stored in the configuration
(Forbes \& Priest 1995; Lin \& Forbes 2000). Charged particles can
be effectively accelerated by the induced reconnection electric
field in the RCS (Martens \& Young 1990; Livenenko \& Somov 1995).

In this paper, we evaluate the induced reconnection electric field
of the X17.2 two-ribbon flare that occurred on 2003 October 28 and
compare it with multi-wavelength and particle observations. Our
results reveal that the reconnection electric field probably plays
an important role in accelerating relativistic nucleons and make a
crucial contribution to the prompt impulsive phase of solar
energetic particles (SEPs).

\section{Observations and data analysis}

\subsection{Flare magnetic reconnection}

The GOES X17.2 two-ribbon flare located in the NOAA active region
10486 (S16E08) began at 09:40 UT, reached its peak at 11:10 UT and
ended around 11:25 UT. Figure 1 shows the RHESSI hard X-ray sources
superimposed on the TRACE 195 {\AA} image. Red contour lines
indicate 12 -- 25 keV thermal bremsstrahlung sources integrated from
11:10:40 to 11:20:40 UT after the flare's peak, and blue ones
indicate 200 -- 300 keV non-thermal sources integrated from 11:06:40
to 11:20:40 UT during the flare's peak and decay phase. It can be
found that the three non-thermal sources are located just on a
flare-post-loop structure, and the thermal ones straddle the arcade
of the loops.

\begin{figure}
\centering
\includegraphics[angle=0,width=8.5cm]{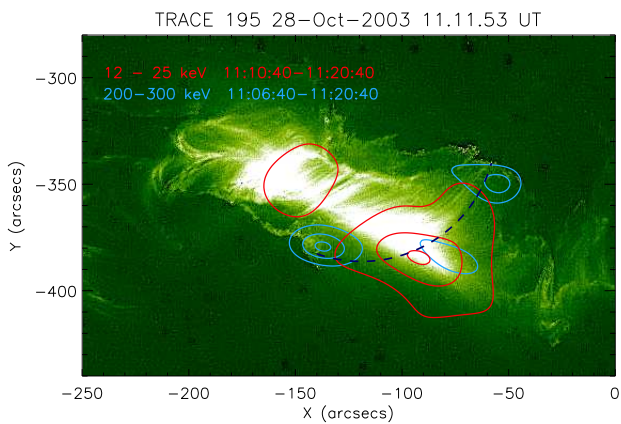}
\caption{RHESSI hard X-ray sources (Rotating modulation collimators
4 -- 8, with MEM-Sato image reconstruction algorithm) overlaid on
the TRACE 195 {\AA} image at 11:11:53 UT. Sky-blue contour lines
indicate 200 -- 300 keV hard X-ray sources integrated from 11:06:40
to 11:20:40 UT, and red 12 -- 25 keV integrated from 11:10:40 to
11:20:40 UT. Dashed blue line indicates the flare-post-loop
structure.} \label{Fig1}
\end{figure}

It is now well-recognized that the flare ribbon's expansion is the
chromospheric signature of the progressive magnetic reconnection in
the corona in which new field lines reconnect at higher and higher
altitudes. The separation motion of flare ribbons that sweep through
the magnetic field lines corresponds to the rate of magnetic
reconnection in the corona, where the reconnection current sheet
(RCS) is generated. We measured the magnetic reconnection rate in
the form of a reconnection electric field, which can be given by
$E_{rec}=VB$ (Forbes \& Lin 2000; Qiu et al. 2002), where V is the
separation velocity of flare ribbons and B the magnetic field that
the ribbons sweep through.

Since the flare occurred near disk center, B can be approximately
taken as the longitudinal component of the magnetic field obtained
from the SOHO/MDI magnetogram at 11:11:03 UT. We use the
high-cadence ($\sim1$ minute) 195 $\rm\AA$ TRACE observations
covering the time interval of 10:47 UT -- 11:26 UT to measure the
flare ribbon's separation. The positions of the flare EUV ribbons
and hard X-ray footpoints were overlaid on the magnetogram (see Fig.
2, detailed processing methods are discussed in Li et al. 2006).
There is no RHESSI data before 11:06:20 UT, so we reconstruct
successive footpoints of 200 -- 300 keV sources between 11:06:20 and
11:10:20 UT (Fig. 2). We can find that the general moving directions
of the two hard X-ray footpoints are toward the south and west and
follow the separation of the flare ribbons. This can be evidence in
support of the model and method for calculating the reconnection
rate.

\begin{figure}
\centering
\includegraphics[angle=0,width=8.5cm]{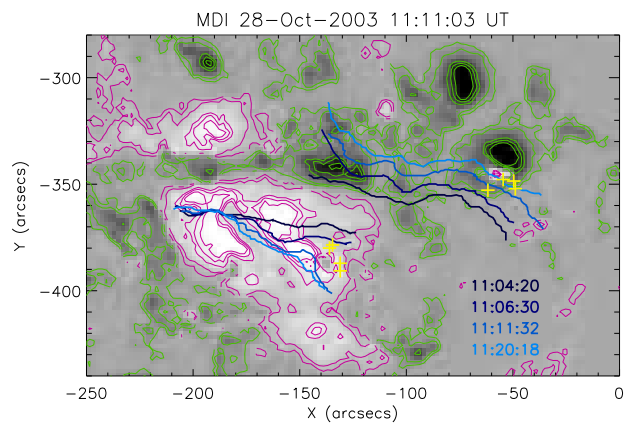}
\caption{Positions of the TRACE EUV ribbons and RHESSI hard X-ray
footpoints overlaid on the magnetogram. Blue lines indicate the EUV
ribbons at different times, and yellow + signs indicate 200 -- 300
keV footpoints for successive integrated time intervals of 100, 130,
180, and 240 s beginning at 11:06:20 UT. Red contours lines indicate
positive longitudinal magnetic field, and green negative.}
\label{Fig2}
\end{figure}

From the MDI magnetogram and the measurement of the flare ribbon's
separation, we get B and V. Then the induced reconnection electric
field can be evaluated. In Fig. 3, the inferred reconnection
electric field is shown in comparison with multi-wave observations
and time profiles of solar neutrons. It is found that the
reconnection electric field has, generally speaking, a good temporal
correlation with hard X-ray and microwave emission, especially with
the neutron capture line and the increase profile of neutron monitor
data. As we know, accelerated protons and nuclei can produce
high-energy neutrons through nuclear reactions from inelastic
collisions in the solar atmosphere, and this process also generates
high energy gamma rays, so the good temporal correlation indicates a
physical link between magnetic reconnection and energy release in
flares, also suggests that the reconnection electric field $E_{rec}$
plays an important role in accelerating nonthermal charged
particles, both electrons and ions.

\begin{figure}
\centering
\includegraphics[angle=0,width=8.5cm]{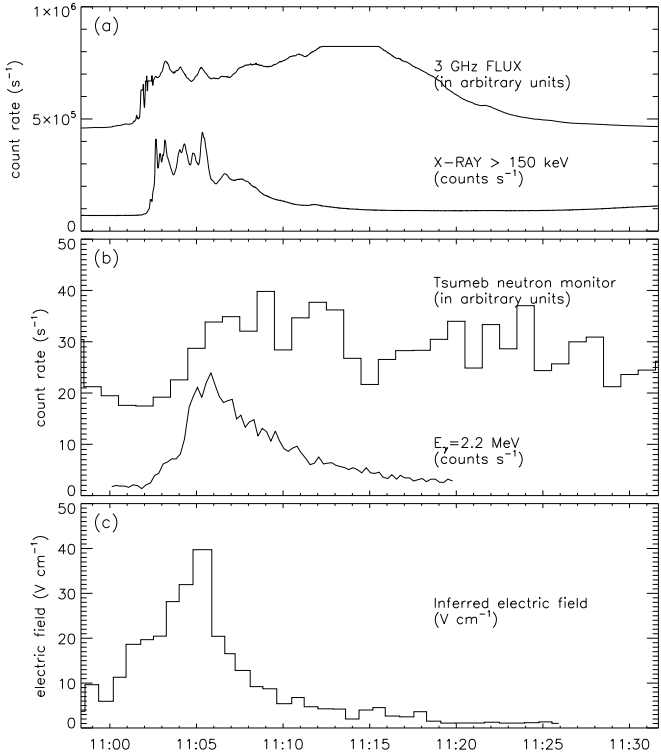}
\caption{(a) Hard X-ray emission (INTEGRAL satellite) and 3 GHz
radio emission (Onde$\check{\rm r}$ejov radio telescope). (b)
Neutron capture line at 2.223 MeV (Kiener et al. 2006, Fig. 2) and
Tsumbe neutron monitor one-minute-averaged count rate. (c)
Reconnection electric field $E_{rec}$ inferred from the X17.2
two-ribbon flare. } \label{Fig3}
\end{figure}

One may suggest different spatial accelerations between electrons
and ions in solar flares. However, INTEGRAL/SPI observations (Kiener
et al. 2006, Figs. 1 and 2) show that the electron-associated hard
X-ray emission and the ion-associated prompt C and O de-excitation
lines (4.4 and 6.6 MeV) have very similar profiles. Comparing the
beginning of the bremsstrahlung emission and the edge of the neutron
capture line increase, a typical neutron thermalization time $\sim$
100 s has been observed (Gros et al. 2004, Fig. 6). Taking this time
delay into account, Hurford et al. (2006) find that the two 2.223
MeV gamma-ray sources have about the same separation as the two
corresponding 200 -- 300 keV electron bremsstrahlung sources with a
displacement of 14 and 17 $\pm$ 5 arcsecs. Thus the electrons and
ions appear to have the same acceleration source. A very simple
explanation of the displacement is that electrons and ions
accelerated in the RCS will travel along different magnetic fields
due to gradient or curvature drift; in other words, electrons are
much more magnetic-controlled than ions.

In the RCS, charged particles can be accelerated by the induced
reconnection electric field. For this event, given the maximum
$E_{rec} \sim 40.0$ V/cm, an acceleration length $l_{acc} \sim
2.5\times10^{7}$ cm is needed to accelerate the protons to GeV
energy. The ratio of the acceleration length to the whole flare
ribbon's length is $\sim 4.0\times10^{-3}$, hence protons are not
accelerated in a single beam running the full length of the RCS.
This avoids the contradiction that the electric current associated
with the accelerated particles would be so strong that the induced
magnetic field would greatly exceed typical coronal values.

\subsection{Solar energetic particles}

During this large solar flare, the near-equatorial neutron monitor
in Tsumeb, with high cutoff rigidity of 9.12 GV, first observed an
enhancement above background before the arrival of solar protons,
which was attributed to direct solar neutrons (Plainaki et al. 2004;
Struminsky 2005). Then a few minutes later, several other stations,
such as Moscow and Apatity neutron monitors, detected excess count
rate of solar cosmic rays (SCRs). A high flux of protons and
electrons were also recorded by GOES and ACE several minutes later.
Figure 4 shows increase profiles of SEPs and SCRs recorded by
satellites and ground-based neutron monitors at 1 AU.

The neutrons emitted at the Sun follow a straight line path from the
emission point to the Earth. According to Tsumeb neutron monitor
data (Fig. 4, panel c), we estimate the neutron's emission time is
11:05 UT $\pm$ 1 minute.

Assuming that protons and electrons travel along the interplanetary
magnetic field (IMF) lines at a speed of $\upsilon$ with no
scattering, in order to compare with multi-wavelength observations,
we estimate the solar release time by subtracting $\Delta t$ from
observed time at 1 AU, where $\Delta t=1.1\,\rm AU/\upsilon-8.3\,\rm
minutes$, and 1.1 AU corresponds to the length of IMF lines when the
solar wind is about 700 km/s (ACE/SWEPAM) for this event. From
ACE/EPAM, GOES-10, and neutron monitors data (Fig. 4, panels a, b,
and d), the evaluated proton's and electron's injection time is
11:12 UT $\pm$ 1 minutes. It is close to the result of Bieber et al.
(2005), who infer the relativistic protons were injected at $\sim$
11:11 UT.

This GLE event was also accompanied by a fast halo CME. Using the
data from the LASCO CME catalog, extrapolating the halo CME to the
solar disk center (quadratic fit), we estimate the upper limit of
the CME liftoff is $\sim$ 11:07 UT, which is later than the
neutron's emission of $\sim$ 11:05 UT. It indicates that the
CME-driven shock plays a minor role in producing relativistic
neutrons, which probably come from the byproducts of charged
particles accelerated in the active region. Hurford et al. (2006)
also suggest that the gamma-ray producing ions appear to be
accelerated by the flare process and not by a widespread shock
driven by a fast CME.

\begin{figure}
\centering
\includegraphics[angle=0,width=8.5cm]{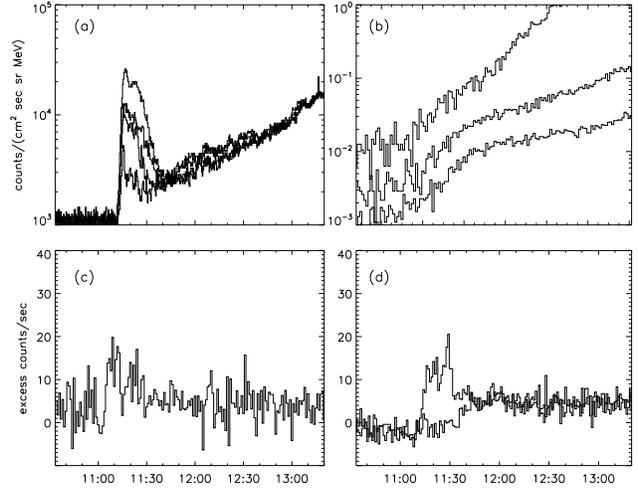}
\caption{(a) The intensity of ACE/EPAM 103 -- 175 keV electron
channel, all four sectors are plotted. (b) GOES-10 $\rm P_{5}$, $\rm
P_{6}$, and $\rm P_{7}$ channel data are 1-minute-averaged and cover
the energy ranges 40 -- 80, 80 -- 165, and 165 -- 500 MeV,
respectively. (c) Tsumbe neutron monitor 1-minute-averaged data. (d)
Apatity and Moscow neutron monitor 1-minute-averaged data.}
\label{Fig4}
\end{figure}

\subsection{Magnetic field configuration}

This event shows another interesting aspect: if relativistic charged
particles (protons and electrons) and their byproduct neutrons
observed at 1 AU are produced almost at the same time in the active
region as we suggest, why is the neutron emission (11:05 UT $\pm$ 1
minute) a few minutes before the proton and electron injection
(11:12 UT $\pm$ 1 minute). To explain this phenomena, we propose the
magnetic-field evolution configuration shown in Fig. 5. This
scenario is based on the model of flares with flux rope ejection
(Ohyama \& Shibata 1998; Lin \& Forbes 2000).

First due to some instability, the coronal flux rope loses
equilibrium and travels upwards, below which the RCS forms. Charged
particles accelerated in the RCS travel downwards along magnetic
field lines and generate microwave emission due to synchrotron, hard
X-ray emission due to bremsstrahlung, gamma-ray emission due to
nuclear reaction, and high energy neutrons in the process of
gamma-ray generation. The produced neutrons follow a straight line
path from the emission point to the Earth undisturbed by magnetic
fields and are detected by the neutron monitors on the ground.
However, the charged particles are probably trapped by the closed
magnetic fields and cannot escape from the active region.

Then a few minutes later (around 11:12 UT), magnetic fields
reconnect at a higher altitude, corresponding to the flare's
expansion in the chromosphere, and the flux rope also travels to a
much higher site and is ejected. During this process, open field
lines should be produced (Shibata 2006), and charged particles
(protons and electrons) accelerated in the RCS could escape along
open field lines into the interplanetary space and be detected at 1
AU.

In fact, during this proposed evolution of the magnetic field
configuration, a large-scale coronal disturbance was observed (Dai
et al. 2005). From the running difference EIT 195 $\rm {\AA}$ images
(shown in Fig. 6), it is found that around 11:12 UT, a large amount
of coronal material was ejected, corresponding to the process from
the coronal brightening to dimming. This process may open quite a
lot of magnetic field lines in the low corona, facilitating the
flare accelerated particles along these open field lines into the
interplanetary space.

\begin{figure}
\centering
\includegraphics[angle=0,width=8.5cm]{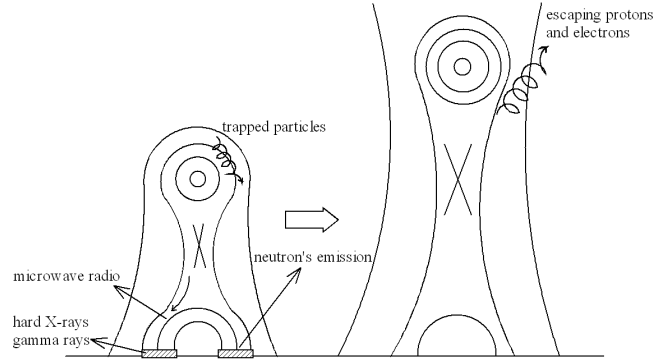}
\caption{Sketch of the proposed model for the magnetic field
configuration associated with the flare and plasmoid ejection. The
subsequent neutron emission, and the proton and electron injection
are shown in the two phases from left to right.} \label{Fig5}
\end{figure}

\begin{figure}
\centering
\includegraphics[angle=0,width=8.5cm]{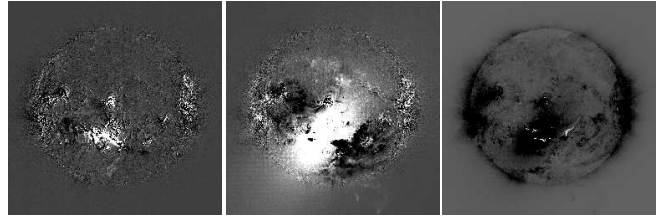}
\caption{The coronal disturbance on 2003 October 28. The three
images show EIT 195 $\rm {\AA}$ at 11:00, 11:12, and 11:24 UT with a
pre-event image subtracted from them.} \label{Fig6}
\end{figure}

\section{Discussion}

The X17.2 flare that occurred on 2003 October 28 in the active
region NOAA 10486 (S16E08) is nominally not well-connected with the
Earth, and this is an argument against flare acceleration of solar
protons. However, because of the high-speed solar wind ($\sim$ 700
km/s), the evaluated footpoint of the IMF is located at $\sim$ W30,
not very far from the flare site. Moreover the large-scale
solar-surface disturbance (shown in Fig. 6) could offer a particle
transport path from the flare site to the well-connected region and
may correspond to a field-opening process. A loop-like IMF line
formed by a preceding CME on 2003 October 26 connecting the Earth
with the flare was also proposed by \cite{Bie05} and \cite{Mir05}.

This GLE event displayed an initial impulsive increase corresponding
to the solar eruption and was followed by a gradual component until
the CME driven shock arrived at the Earth, as Fig. 7 shows. It
appears that there are two populations of SEPs: the prompt one
causes an impulse-like increase and the delayed one has a slow
intensity rise, especially for the lower energetic particles, as
shown in the middle two panels. The prompt solar energetic particles
could be well-explained by the acceleration in the flare active
region; however, the following delayed ones cannot stem from only
the active region because of their long-duration injection. Some of
the particles accelerated in the active region would be trapped or
reaccelerated by the coronal CME-driven shock, and the
interplanetary CME-driven shock could also add some particles to the
delayed injection.

\begin{figure}
\centering
\includegraphics[angle=0,width=7.5cm]{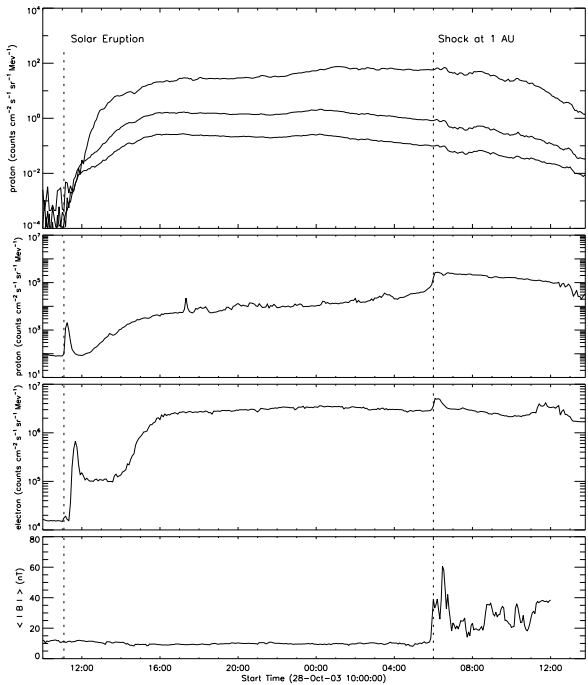}
\caption{Temporal profiles of the energetic protons, electrons
intensity, and magnetic field near 1 AU of the 2003 October 28
event. In the upper panel, the GOES-10 $\rm P_{5}$, $\rm P_{6}$, and
$\rm P_{7}$ channel data cover the energy ranges 40 -- 80, 80 --
165, and 165 -- 500 MeV, respectively. Lower energetic protons (0.31
-- 0.58 MeV) and electrons (0.038 -- 0.053 MeV) from ACE/EPAM are
shown in the middle two panels and magnetic field from ACE/MAG at
bottom.} \label{Fig7}
\end{figure}

From Fig. 7, it is also found that, when the CME-driven shock
reached 1 AU at $\sim$ 6:00 UT Oct 29, the lower energetic particles
(ACE/EPAM 0.31 -- 0.58 MeV protons and 0.038 -- 0.053 MeV electrons)
showed obvious increase. However, the relatively much higher
energetic particles (GOES several tens of MeV protons) show nearly
no increase. This indicates that CME-driven shocks play a minor role
in accelerating higher energetic, especially the relativistic
particles.

The GLE event of 2003 October 28 presents many unusual features all
of which we do not explain. From the analysis of multi-wavelength
observations and energetic particle data, we propose that the flare
magnetic reconnection, especially the induced electric field, makes
a crucial contribution to the prompt relativistic particles. On the
other hand, the CME-driven shock probably plays an important role in
the delayed gradual injection of SEPs, especially in the lower
energetic ones.

\begin{acknowledgements} We are very grateful to the referee Dr. E.
Kontar, whose constructive comments have greatly improved this
paper. We thank the TRACE, SOHO, RHESSI, and ACE teams for providing
the observational data. The INTEGRAL satellite and Onde$\check{\rm
r}$ejov radio telescope data were kindly provided by Dr. M.
Karlick$\acute{\rm y}$. We thank the Polar Geophysical Institute
(Russia) and Unit of Space Physics (South Africa) for providing the
neutron monitor data. This work was supported by NSFC key projects
No. 10333040, No. 10221001, and NKBRSF of China G2006CB806300.
\end{acknowledgements}

\end{document}